\documentclass[]{spie}  %>>> use for US letter paper
\usepackage{amsmath,amsfonts,amssymb}
\usepackage{graphicx} 
\usepackage{lipsum} 
\usepackage[colorlinks=true, allcolors=blue]
{hyperref}

\title{Performance estimation of photonic integrated wavefront corrector for single-mode fiber coupling}

\author[a]{Dhwanil Patel}
\author[a]{Momen Diab}
\author[b]{Ross Cheriton}
\author[a,c]{Jacob Taylor}
\author[a]{Libertad Rojas}
\author[a,c]{Suresh Sivanandam}

\affil[a]{Dunlap Institute for Astronomy and Astrophysics, University of Toronto, Toronto, Ontario, Canada}
\affil[b]{Quantum and Nanotechnologies Research Centre, National Research Council Canada, Ottawa, Ontario, Canada}
\affil[c]{David A. Dunlap Department of Astronomy and Astrophysics, University of Toronto, Toronto, Ontario, Canada}

\authorinfo{Further author information: (Send correspondence to Momen Diab)\\Momen Diab: E-mail: momen.diab@utoronto.ca}
\begin{document} 
\maketitle

\begin{abstract}
%\lipsum[1-1]
Many modern astronomical instruments rely on the optimal coupling of starlight into single-mode fibers (SMFs). For ground-based telescopes, this coupling is limited by atmospheric turbulence. We propose an integrated wavefront corrector based on silicon-on-insulator (SOI) photonics, which samples the aberrated wavefront via a microlens array (MLA). The MLA focuses the sampled wavefront onto an array of grating couplers that inject the beamlets into the single-mode waveguides of the corrector. The beams in each waveguide are then shifted in phase using thermo-optic phase shifters before combining the co-phased beams into one single-mode waveguide. %The mode of the combined beam is then expanded to match the mode of the single-mode fiber sitting at the edge of the corrector for optimal outcoupling. 
In this work, we analyze the external factors that we anticipate will impact the performance of the corrector. Specifically, we study the effects of the telescope pupil function with obscuration, determine whether the corrector requires tip/tilt pre-correction, and analyze the impact of scintillation on the correction quality.
\end{abstract}


\section{Introduction}
\label{sec:intro}  % \label{} allows reference to this section
%Efficiently coupling starlight into single-mode fibre (SMF) has been a limiting factor for many fiber-fed astronomical instruments with applications ranging from interferometry based on beam combiners \cite{}, spectroscopy using arrayed waveguide gratings\cite{}, OH suppression\cite{} and many other applications \cite{}. The primary obstacle to optimal coupling is the aberration due to the atmosphere. For optimal coupling, the profile of starlight should match that of the SMF; a fundamental Gaussian mode with a flat wavefront. But since the starlight propagates through the atmosphere, wavefront errors are introduced with mode deviating from the fundamental mode. To overcome this, an adaptive optics (AO) system can be used to reduce the wavefront error with on-sky demonstrations at the Subaru telescope and Large Binocular Telescope reaching the coupling efficiency of $> 70\%$ \cite{jovanovic2017efficient}, \cite{bechter2019characterization}. 

Coupling starlight into single-mode fibers (SMFs) is a limitation for many fiber-fed astronomical instruments, with applications ranging from interferometry based on photonic beam combiners \cite{minardi2019discrete, vievard2020first}, spectroscopy using arrayed waveguide gratings \cite{crass2019need, gatkine2019astrophotonic, stoll2017high}, OH suppression fiber filters \cite{ellis2012suppression}, and many others \cite{ellis2021general, jovanovic20232023}. The primary obstacle to efficient coupling is atmospheric aberration. Earth's atmosphere causes the Airy disk at the telescope focus to break into a fast-evolving speckle pattern that does not match the Gaussian-like mode of the SMF. An adaptive optics (AO) system can be used to correct the phase errors in the wavefront and enhance the coupling efficiency. %For effective coupling, the profile of starlight should match that of the SMF: a fundamental Gaussian mode with a flat wavefront. However, as starlight propagates through the atmosphere, wavefront errors are introduced, causing the mode to deviate from the fundamental mode. To mitigate this, an adaptive optics (AO) system can be used to reduce the wavefront error. 
On-sky demonstrations at the Subaru Telescope \cite{jovanovic2017efficient} %and the Large Binocular Telescope (LBT) \cite{bechter2019characterization} 
have achieved coupling efficiencies $>50\%$ using a complex extreme AO (ExAO) system that compensates for most of the distortion in the wavefront.

Inefficiencies in AO-assisted coupling arise because the correction is typically partial, with the number of corrected modes not meeting the requirements for perfectly flattening the wavefront. This partial correction occurs because the AO system is limited by errors in its three components: the wavefront sensor (WFS), the real-time controller (RTC), and the wavefront corrector (WFC). When the WFC is a deformable mirror (DM), the inability to exactly and instantaneously achieve the commanded shape adds a fitting and a temporal error, respectively, reducing the correction quality. Additionally, the WFC must meet certain stroke and pitch requirements that the two main current DM technologies --- voice coils and microelectromechanical systems (MEMS) --- can only partially fulfill, often compromising one aspect to satisfy the others. The high cost of DMs also precludes instrument concepts that require a DM for each object in a crowded starfield, e.g., multi-object AO (MOAO) and multi-object spectroscopy (MOS).
%The inability of commonly used pupil-plane wavefront sensors (WFS) to detect petal modes,  blind modes, and other non-common path aberrations. These issues can be mitigated by using photonics-based focal-plane WFS such as photonic lanterns (PL) \cite{diab2019modal, norris2020all}. Due to the ability of PL to be mode-selective, hybrid-PL can be used both for wavefront sensing and optimal coupling of starlight into SMF, either with or without partial lower-mode correction with AO \cite{diab2021starlight} \cite{norris2022optimal}. These solutions still rely on using deformable mirrors (DM) for correcting the abberated wavefront and while a DM can achieve a higher mechanical stroke, it tends to be expensive and challenging to multiplex efficiently. 

\begin{figure}[hbt!]
    \centering
    \includegraphics[width=\textwidth]{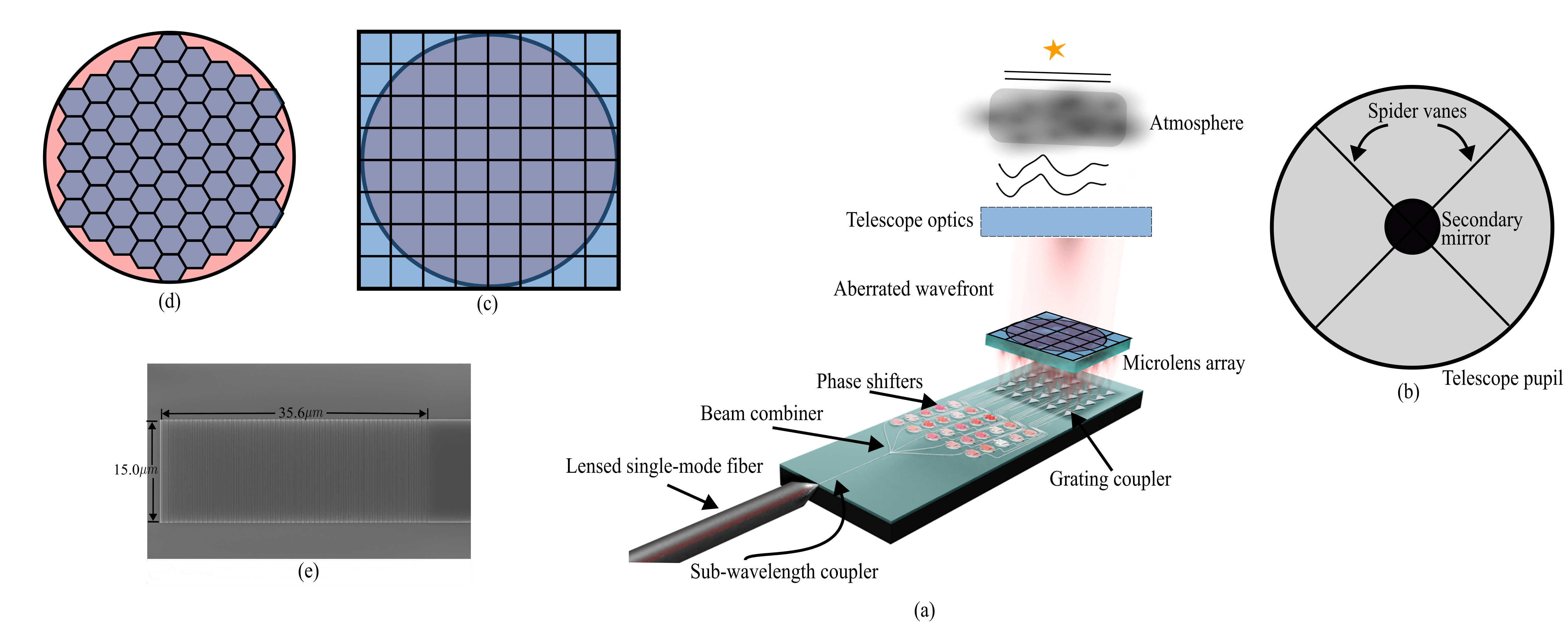}
    \caption{The schematic of the integrated phase corrector in (a) with the MLA that samples the light from the obscured telescope pupil (shown in (b)). Two different geometries of the MLA with $8\times 8$ squares and $61$ hexagons are shown in (c) and (d), respectively. (e) A scanning electron micrograph of one grating coupler.}
    \label{fig: picSchematic}
\end{figure}

%To mitigate the same issues with SMF coupling and not rely on DM, 
We propose a photonic WFC to couple starlight distorted by the atmosphere into SMFs efficiently. The integrated device shown in Fig. \ref{fig: picSchematic}a can be multiplexed to enable application in fiber-based MOS for $\sim 100$s of objects. The photonic integrated chip (PIC) uses a microlens array (MLA) bonded to the PIC to spatially sample the aberrated wavefront from the telescope's exit pupil. The MLA focuses the wavefront onto grating couplers optimized for vertical coupling into single-mode waveguides in the PIC. The phase of the coupled light in each of the spatial channels is modulated using thermo-optic phase shifters. The co-phased beams are then coherently combined into a single-mode waveguide using a beam combiner.% The mode of the combined light is then matched to that of an SMF using a sub-wavelength grating (SWG) waveguide \cite{cheben2015broadband} for optimal outcoupling into an SMF. 

We showed a proof-of-concept simulation with an idealized model of the PIC components in \cite{diab2022photonic}. We also performed more extensive end-to-end simulations of the PIC with components that match the designs of the fabricated PIC in the laboratory to gauge the total throughput in \cite{Patel:24}. In this work, we estimate the performance of the PIC considering external artifacts that affect the system. We analyze the effect of the telescope pupil with a secondary obscuration and spider vanes. We also study whether the use of a fast-steering mirror (FSM) is required upstream of the PIC. Finally, we analyze the effects of scintillation and formulate an analytical solution for the performance metric of the PIC. The methods and simulation tools used are described in Sec. \ref{sec: methods}. In Sec. \ref{sec: pupilplaneMLA}, the simulation results for telescope pupils with various obscuration ratios are given. In Sec. \ref{sec: FSM}, the advantage of tip/tilt pre-correction on the performance of the PIC at different turbulence strengths is investigated. Finally, we show how the correction quality of the PIC is affected by the effects of scintillation in Sec. \ref{sec: scinti}. The results are discussed in Sec. \ref{sec: discussion} with a conclusion at the end in Sec. \ref{sec: concl}. 

\section{Simulation methods}\label{sec: methods}
To assess the performance of the PIC, we perform simulations that begin with the generation of phase screens that have a von Kármán power spectral density\cite{welsh1997fourier}. The phase screens simulate the distorted wavefront after propagating through turbulent atmospheric layers at specified altitudes. The turbulence strength ($D/r_0$) is governed by the Fried parameter ($r_0$) and the diameter of the telescope pupil ($D$).
%The spectral density is governed by Fried's parameter $r_0$, the spatial frequency $f_s$, and the outer scale $L_0$. 
We perform simulations over a range of different Fried parameters to determine the correction quality for a given telescope diameter used in the simulations.

The generated phase screens are propagated through a free space distance of $50$ km along the line of sight to the telescope using the Fresnel diffraction integral. The long propagation distance is the worst-case scenario for the distance from the tropopause for targets at low elevation angles ($10$ deg). This introduces scintillation effects and causes intensity fluctuation in the received wavefronts. To minimize aliasing due to Fresnel propagation, the generated screens are $3$ times larger than the aperture of the telescope. The telescope's aperture function with secondary obscuration and spider vanes is applied to the complex field. The field at the telescope pupil is then demagnified and imaged onto the MLA with square and hexagonal configurations (see Fig. \ref{fig: picSchematic}c and d). The corresponding focal spots at the focal plane of each lenslet of the MLA are calculated using the Fraunhofer diffraction integral. The focal spots are then line-scanned to get a 1D field and propagated through a 2D model of the grating couplers using a finite-difference time-domain (FDTD) solver. The coupled fields in each channel are co-phased and combined using a model of a multimode interferometer (MMI) beam combiner \cite{hosseini2009output}. The beam propagation method (BPM) is used to model the MMI model. Finally, we simulate the propagation of the combined field through a sub-wavelength grating waveguide \cite{cheben2015broadband} that expands the mode to match that of an SMF for an optimal coupling from the PIC to the SMF using FDTD. 

To quantify the performance of the photonic WFC, we adopt the concept of the Strehl ratio. Since the total power at the output of the SMF is correlated to the quality of the point spread function (PSF) \cite{diab2021starlight} \cite{faucherre2000using}, we define the photonic Strehl ratio ($\mathit{SR}_{\mathit{ph}}$) as the ratio of the total output SMF power at the seeing-limited condition to the total SMF power at the diffraction limit. Note that this metric provides the inherent performance of the PIC and does not account for the insertion loss from free space into the grating couplers, the propagation loss through the PIC components, and the coupling loss from the PIC to the SMF. These losses are characterized in \cite{Patel:24}.

\section{Pupil plane of telescope} \label{sec: pupilplaneMLA}
\begin{figure}[hbt!]
    \centering
   \includegraphics[width=\textwidth]{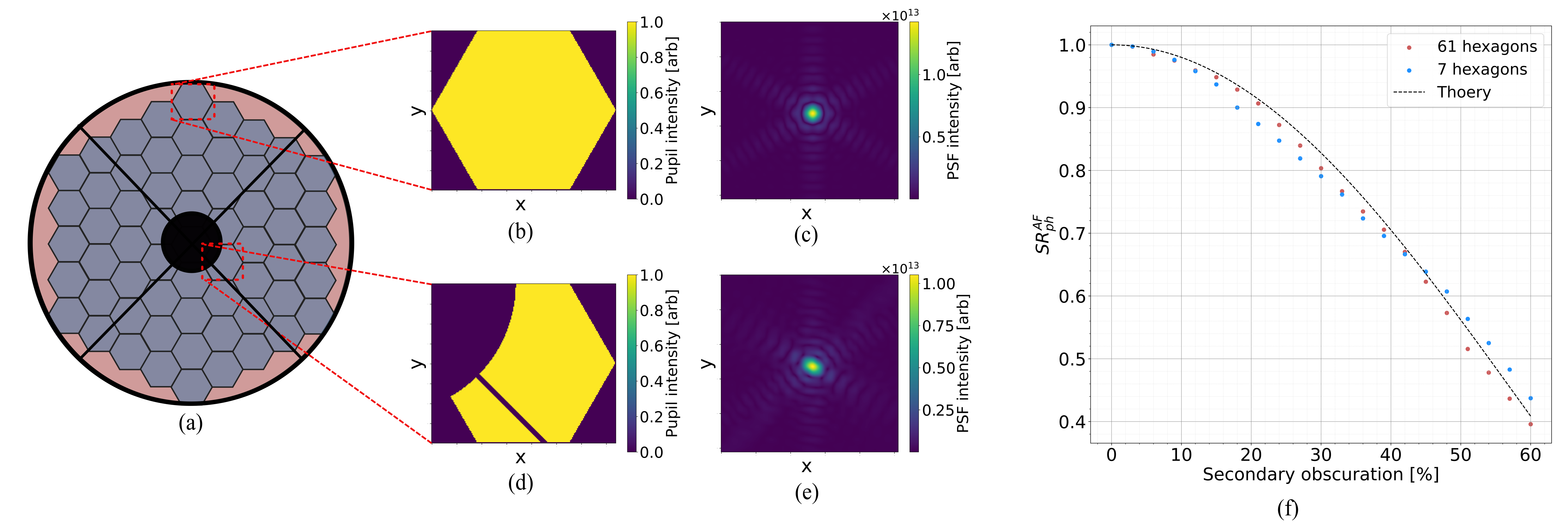}
    \caption{(a) The demagnified image of the telescope aperture function with a $20\%$ secondary obscuration and spider vanes overlaid on the MLA with $61$ hexagonal lenslets. An unobscured and a partially obscured subapertures of the pupil are shown in (b) and (d), respectively. The corresponding focal patterns are shown in (c) and (e), respectively. The PSF degradation $\mathit{SR}_{\mathit{ph}}^{\mathit{AF}}$ as a function of secondary obscuration is shown in (f) for the two MLA configurations of $7$ and $61$ hexagons. The plot also compares the two curves with the theoretical prediction given by Eq. \ref{eq: COT}.}
    \label{fig: telescope_pupil_psf}
\end{figure}
The first simulation aims to study the effect of the secondary mirror and its spider vanes on the performance of the PIC (Fig. \ref{fig: telescope_pupil_psf}a).
%We want to analyze how the PIC performs when the telescope's pupil is obstructed by the secondary mirror and its support structures. The telescope's pupil function, therefore, includes a secondary obstruction and spiders, as illustrated in Figure \ref{fig: telescope_pupil_psf}(a). 
The effect of the secondary obscuration on the PSF of a full circular pupil is to redistribute light away from the Airy disk to the diffraction rings. As the obscuration ratio increases, more optical power is transferred from the disk to the rings. The spider vanes introduce diffraction spikes, the geometry of which depends on the arrangement and the orientation of the vanes. When the pupil is spatially sampled by an MLA, some subapertures are not fully illuminated, and their respective PSFs are distorted. This adversely affects the coupling efficiency of a subset of the grating couplers.

%The PSF from such an obstructed telescope pupil function has been studied extensively.
For the unsampled pupil and the individual subapertures, the degradation of the PSF due to secondary obscuration and spider vanes can be quantified using a modified definition of the Strehl ratio where the peak intensity of the diffraction-limited PSF of the obscured pupil is divided by the peak intensity of the diffraction-limited PSF of the clear pupil\cite{harvey1995diffraction}. This modified aperture function Strehl ratio $\mathit{SR}^{\mathit{AF}}$ is equal to the squared area ratio of the unobscured and clear pupil functions according to the central ordinate theorem\cite{harvey1995diffraction}\cite{gaskill1978linear}:

%comparing the imaginary amplitude spread functions derived from the complex pupil functions of both obscured and unobscured systems. This degradation is expressed as the squared ratio of these amplitude spread functions, corresponding to the ratio of the degraded PSF to the ideal diffraction-limited PSF \cite{harvey1995diffraction} \cite{gaskill1978linear}. Thus, we can theoretically quantify the PSF degradation due to obscuration using:

\begin{equation}
    \label{eq: COT}
    \mathit{SR}_{\mathit{ph}}^{\mathit{AF}} \approx
    \mathit{SR}^{\mathit{AF}} = \frac{\left(\text{Area}_{\text{  unobscured pupil}} - \left(\text{Area}_{\text{  secondary}} + \text{Area}_{\text{  spiders}} \right)\right)^2}{\left({\text{Area}_{\text{  unobscured pupil}}}\right)^2},
\end{equation}
where the ratio $\mathit{SR}_{\mathit{ph}}^{\mathit{AF}}$ is defined as the total power at the output SMF with an obscured pupil to the total power with an unobscured pupil, analogous to the photonic Strehl ratio defined earlier.

Since we use an MLA and sample the pupil into sub-apertures, not all subapertures are obstructed (see Fig. \ref{fig: telescope_pupil_psf}a). Fig. \ref{fig: telescope_pupil_psf}b and Fig. \ref{fig: telescope_pupil_psf}d show two such examples of pupil functions, with no obstruction and with obstruction, respectively. The PSFs of the corresponding pupil functions are shown in Fig. \ref{fig: telescope_pupil_psf}c and \ref{fig: telescope_pupil_psf}e for no obstruction and with obstruction cases, respectively. We simulated pupil functions at the diffraction limit with obscuration ratios up to $60\%$  and spider vanes with widths that are $2.5\%$ of the secondary radius. The sampled subapertures are then propagated through the PIC as described in Sec. \ref{sec: methods} to get the total power in the output SMF. The corresponding $\mathit{SR}_{\mathit{ph}}^{\mathit{AF}}$ are then calculated by normalizing the power for each obscured case by the power of the unobscured case. The photonic $\mathit{SR}_{\mathit{ph}}^{\mathit{AF}}$ as a function of obstruction ratio is shown in Fig. \ref{fig: telescope_pupil_psf}f. 

\section{Tip/tilt correction} \label{sec: FSM}
\begin{figure}[hbt!]
    \centering
        \includegraphics[width=\linewidth]{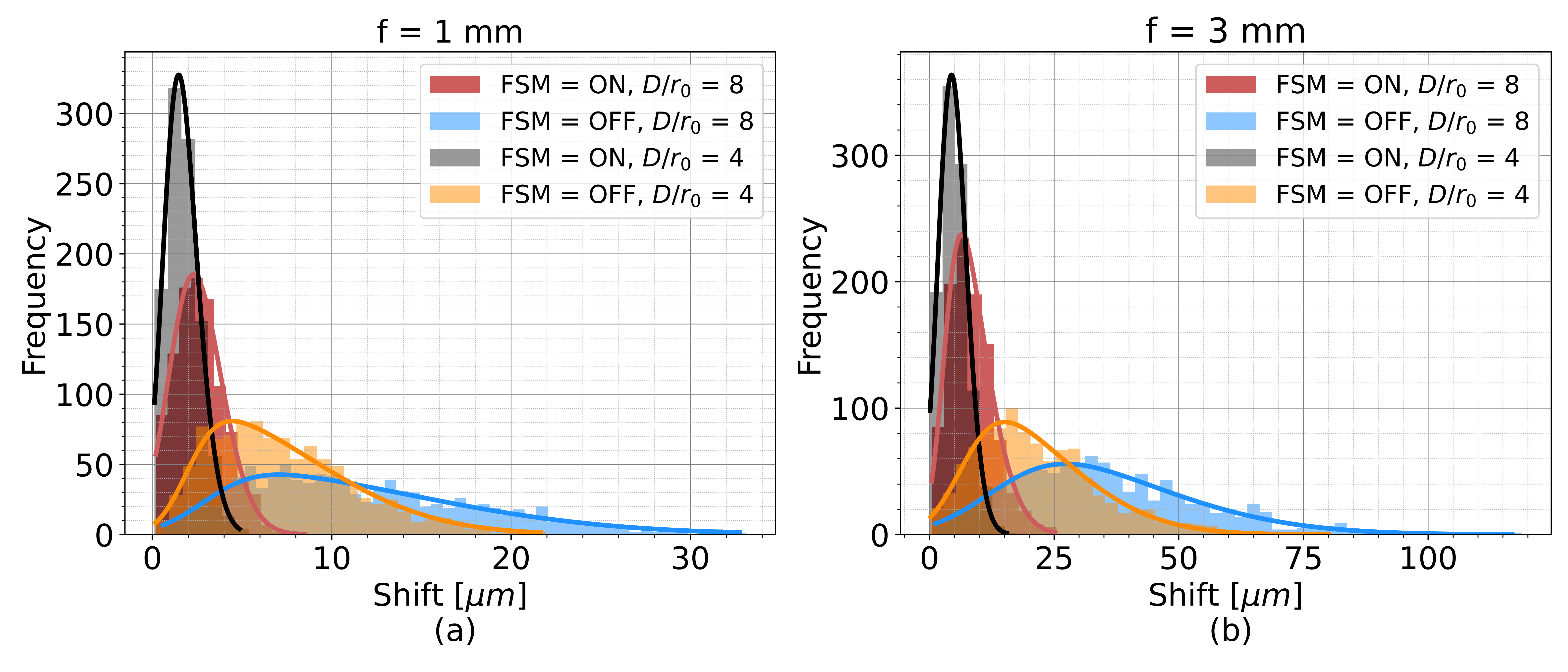}
    \caption{Histogram of the shifts in the MLA focal spots with focal lengths (a) $f=1$ mm and (b) $f=3$ mm. The different curves compare the cases with and without an upstream tip/tilt pre-correction for turbulence strength $D/r_0 = 4$ and $8$.}
    \label{fig: focalspot_dist}
\end{figure}

Tip/tilt errors in the wavefront cause shifts in the focal spots of the MLA from their optimal coupling positions on the grating couplers. To minimize this effect, we assumed lenslets with short focal lengths. We compared the performance of MLAs with focal lengths of $1$ mm and $3$ mm. Specifically, we sampled focal spots generated from 1000 phase screens at the worst-case turbulence scenario in our simulation pipeline, i.e., $D/r_0 = 8$. The centroid of each focal spot was calculated to estimate the shift caused by the tip/tilt errors. As shown in Fig. \ref{fig: focalspot_dist}a, approximately one-third of the focal points shifted less than $7.5$ $\mu$m with a focal length $1$ mm, compared to less than $3\%$ with the $3$ mm MLA as shown in Fig. \ref{fig: focalspot_dist}b. Focal spots shifting less than $7.5$ $\mu$m fall within the active region of the grating couplers, which span $15$ $\mu$m at their shorter dimension (see Fig. \ref{fig: picSchematic}e). %While the grating coupler model used has an optimal coupling position off-center as shown in \cite{Patel:24}, we are more interested in the correction ability of the PIC. 

\begin{figure}[hbt!]
    \centering
    \includegraphics[width=0.8\textwidth]{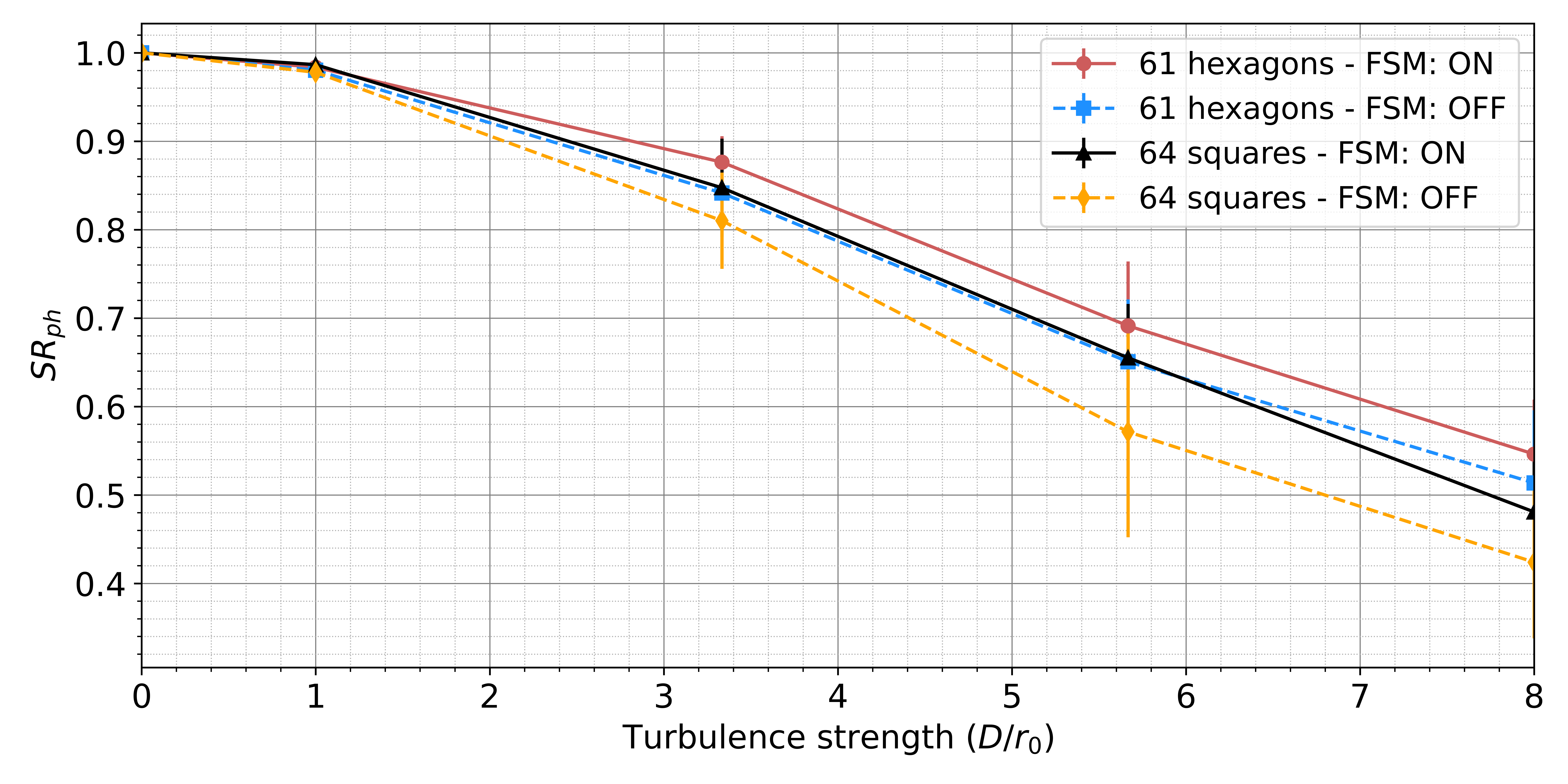}
    \caption{Photonics Strehl ratio decreases with turbulence strength. The solid red and dashed blue curves compare the PIC correction performance with $61$ hexagonal MLA with and without tip/tilt correction, respectively. The solid black and dashed yellow lines are for the $8\times 8$ square MLA arrangement.}
    \label{fig: FSM_Dr0vsSR}
\end{figure}

We also quantified the advantage of using an upstream FSM to correct the tip/tilt errors before coupling the wavefront in the PIC. The ideal FSM is simulated by mathematically removing the tip and tilt modes from the incoming wavefront before sampling with the MLA. 
%With some of the spots remaining within the grating coupler surface with $1\,mm$ MLA, we then compared the correction ability of PIC with and without lower-order mode correction from FSM in the upstream. To simulate the FSM, we mathematically remove the lower-order modes from the generated wavefront before sampling wavefront subapertures through the MLA. This is followed by the steps of the simulation pipeline explained in section \ref{sec: methods}. 
Fig. \ref{fig: FSM_Dr0vsSR} compares the correction performance of the PIC over a range of turbulence strength for $61$ hexagonal and $8\times 8$ square arrangements of MLA.

\section{Scintillation effects} \label{sec: scinti}
\begin{figure}[hbt!]
    \centering
    \includegraphics[width=\textwidth]{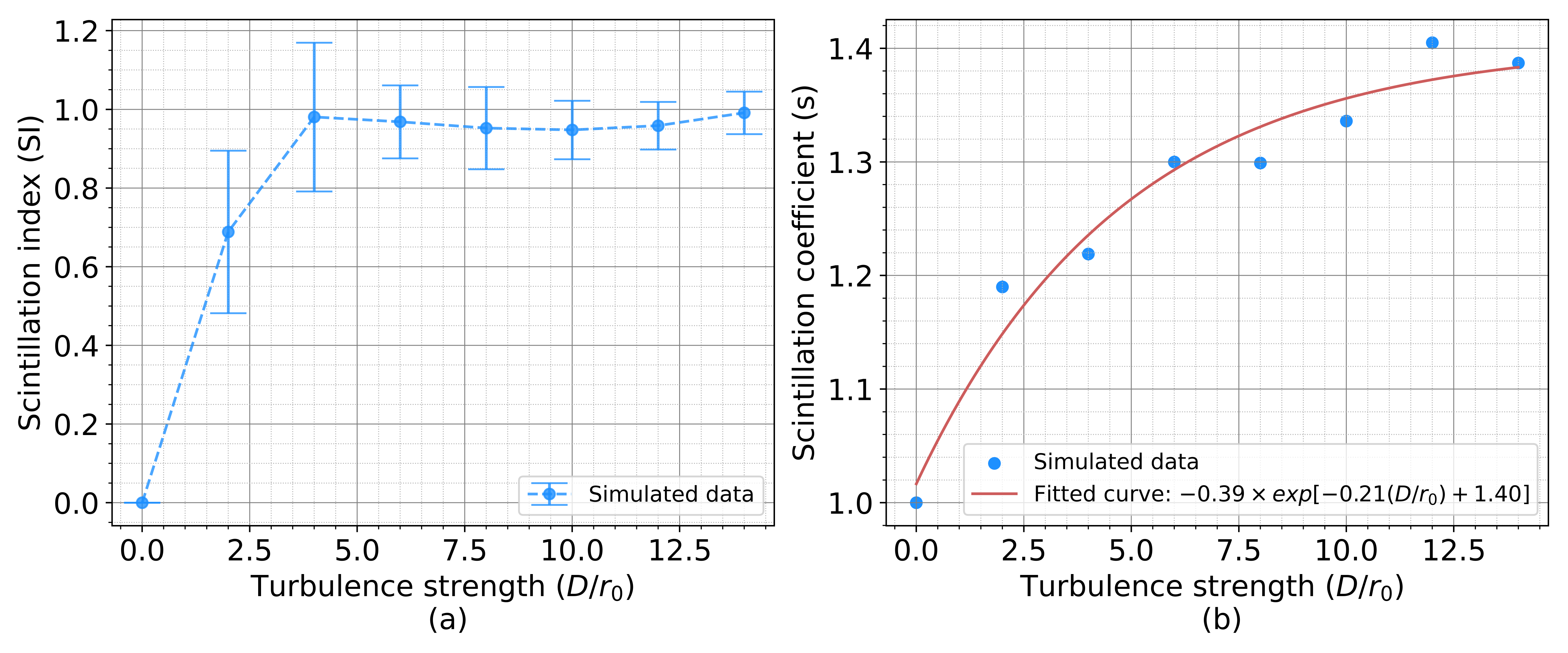}
    \caption{(a) Scintillation index ($\mathit{SI}$) as a function of the turbulence strength for the integrated wavefront corrector with a 61 hexagons MLA. (b) For the same arrangement, the scintillation coefficient ($s$) as a function of $D/r_0$ with an exponential function fitted.}
    \label{fig: SI_SC_Dr0}
\end{figure}

Next, we examined how intensity fluctuations in the wavefront impact the PIC's correction ability. These intensity fluctuations are characterized by the scintillation index, defined as the relative variance in the pupil's intensity: 
\begin{equation}
    \label{eq: SI}
    \mathit{SI} = \frac{\left<I^2\right> - \left<I\right>^2}{\left<I\right>^2},
\end{equation}
where $I$ is the intensity distribution of the wavefront at the telescope pupil. Figure \ref{fig: SI_SC_Dr0}a shows the scintillation index over various turbulence strengths.

For DMs, the mirror ideally assumes the inverse shape of the wavefront to correct aberrations. This correction ability is limited by the number of degrees of freedom, or actuators, that the mirror has. Hence, DMs are characterized by the fitting error in the corrected wavefront. For a DM with $M$ segments along its longest dimension, the fitting error is given by \cite{hudgin1977wave}:
\begin{equation}
    \label{eq: fiterr_DM}
    \sigma^2 = \alpha \left(\frac{1}{M}\frac{D}{r_0}\right)^{5/3},
\end{equation}
where $\alpha = 1.26$ for a piston-only segmented DM.
We showed in \cite{Patel:24} an equivalent fitting error for the photonic wavefront corrector. The Strehl ratio falls exponentially with wavefront error ($\mathit{SR} \approx \exp\left(-\sigma^2\right)$), and thus the photonic Strehl ($\mathit{SR}_{\mathit{ph}}$) of a photonic wavefront corrector can be approximated by:
\begin{equation}
    \label{eq: SR_ph_fiterr}
    \mathit{SR}_{\mathit{ph}} \approx \exp\left[-s\cdot\alpha\left(\frac{1}{M}\frac{D}{r_0}\right)^{\beta}\right],
\end{equation}
where the fitting error coefficient $\alpha $ depends on the geometry of the MLA. % and is 0.73 and 0.86 for the hexagonal and square arrangements respectively. 
 The exponent of the fitting error $\beta$ is equal to $0.85$ in the case of the photonic corrector\cite{Patel:24}. The scintillation coefficient ($s$) encapsulates the effect of scintillation on the quality of the correction. %is encapsulated by the scintillation coefficient $s$. %($M$) is the dimension of the MLA array along the longest dimension (e.g., in the case of 61 hexagons, $M=9$)

\begin{figure}[hbt!]
    \centering
    \includegraphics[width=0.8\textwidth]{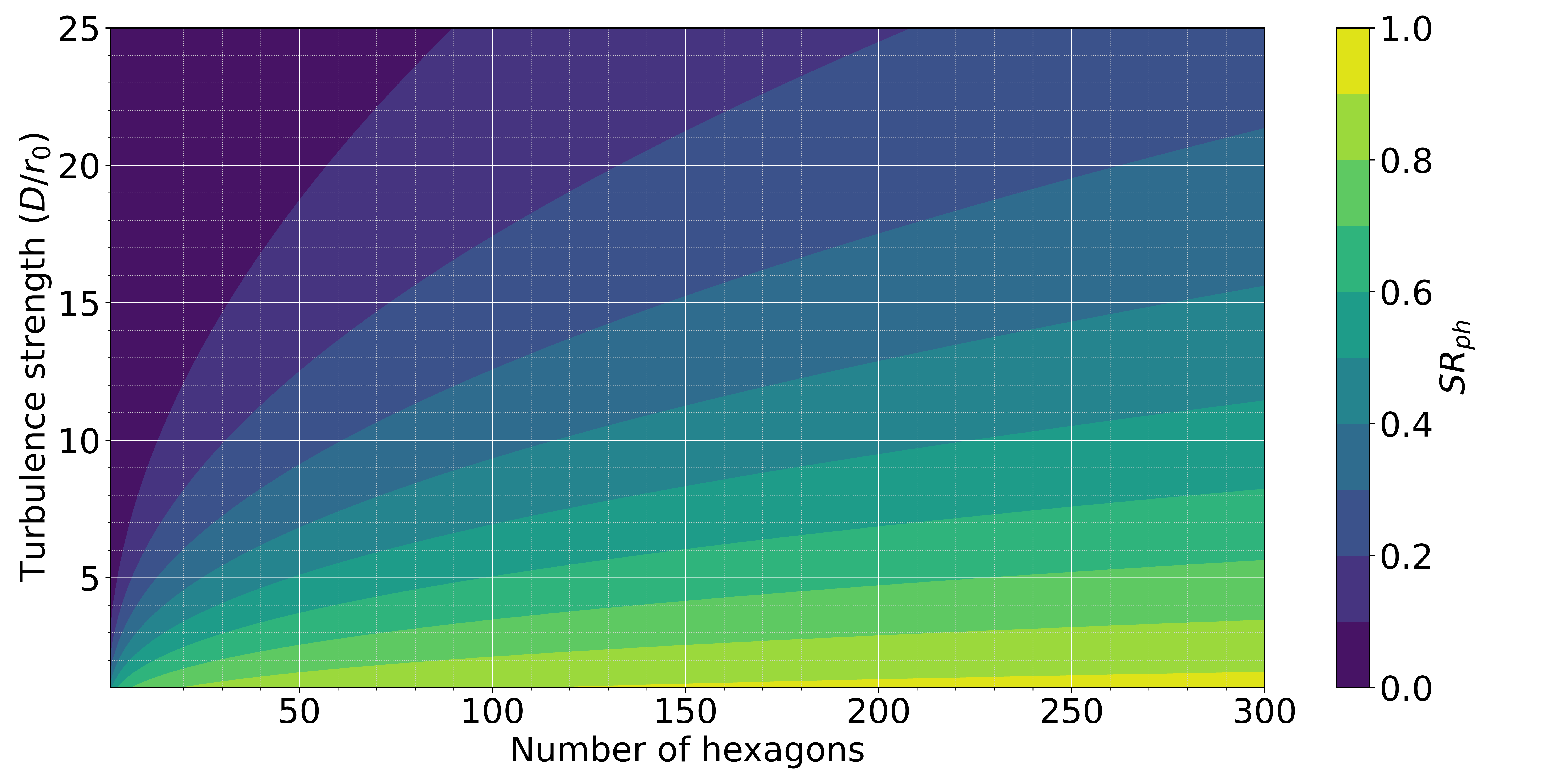}
    \caption{The dependence of the photonic Strehl ratio on the turbulence strength and the number of subapertures.}
    \label{fig: Dr0vsnumHexvsP}
\end{figure}

Taking the distance to the highest turbulence layer to be $50$ km (at the lowest elevation angle of $10$ deg), we calculated the photonic Strehl ratio for turbulence strengths from the diffraction limit to  $D/r_0=14$. Fitting Eq. \ref{eq: SR_ph_fiterr}, we established the relationship between the scintillation coefficient ($s$) and the turbulence strength ($D/r_0$), as shown in Fig. \ref{fig: SI_SC_Dr0}b. This allows us to derive an empirical relation between $\mathit{SR}_{\mathit{ph}}$ and turbulence strength for a given MLA arrangement with dimension $M$. Using the empirical relation, we can predict the performance of the wavefront corrector based on its size as shown in Fig. \ref{fig: Dr0vsnumHexvsP}.

\section{Discussion} \label{sec: discussion}
%- how is performance affected by obscuration. How is it worse/better in the case of PAOWER compared to classical DMs.
%- Importance of tip/tilt correction. why centroid graphs contradict SR graphs.
%- Significance of scinitillation. What is different here compared to the OpEx paper.
%We studied the effects of secondary obscuration on the performance of the WFC for configurations of 61 and 7 hexagons. 
The Strehl ratio ($\mathit{SR}$) is chosen as a performance metric for the photonic WFC. In imaging systems, the $\mathit{SR}$ is defined as the ratio of the peak intensity of the aberrated PSF to the peak intensity of the diffraction-limited PSF. We define two modified Strehl ratios for the photonic phase corrector. The photonic Strehl ratio ($\mathit{SR}_{\mathit{ph}}$) is the ratio of the power in the PIC's output SMF to the power in the output SMF under diffraction-limited conditions, i.e., when a planar wavefront is incident on the MLA. The second Strehl ratio, introduced in Sec. \ref{sec: pupilplaneMLA}, quantifies the effect of pupil obscuration. The aperture function Strehl ratio for the photonic WFC ($\mathit{SR}_{\mathit{ph}}^{\mathit{AF}}$) is a modification on $\mathit{SR}^{\mathit{AF}}$ \cite{gaskill1978linear}, similar to the modification made to define $\mathit{SR}_{\mathit{ph}}$ from $\mathit{SR}$.   

The secondary obscuration degrades the sampled PSF and affects the coupling efficiency of the grating couplers behind it. %To determine whether these effects further impact the correction quality of the WFC. 
As shown in Fig. \ref{fig: telescope_pupil_psf}f, $\mathit{SR}_{\mathit{ph}}^{\mathit{AF}}$ for both configurations decreases as the obscuration ratio increases, similar to $\mathit{SR}^\mathit{AF}$. %according to the theory in equation \ref{eq: COT}. 
This indicates that the power at the output SMF of the photonic WFC does not experience additional degradation beyond that caused by the obscuration on the PSF of a full pupil. Consequently, no further mitigation is expected to be necessary for the correction quality, except for the exclusion of insufficiently illuminated subpertures. This is analogous to conventional AO systems, where certain actuators are either floated or made to follow their neighbors when controlled zonally because their corresponding WFS subapertures do not receive enough light. A grating coupler array tailored to a specific aperture function would only include gratings where enough light is focused on the PIC. In this way, the dark subapertures would not adversely affect the coherent combination scheme.%and an optimal secondary obscuration configuration must be found to minimize the independent segmented piston modes of the DM that the WFS cannot detect \cite{clare2021prism}. 

To a first approximation, the tip/tilt effect could be mitigated by designing the grating couplers to handle fast beams. A short focal-length MLA would minimize the shift in the focal spots caused by tilt in the pupil plane. Additionally, an FSM could be used to pre-correct the collected wavefront before coupling. With a $1$ mm focal-length MLA, the gain in $\mathit{SR}_{\mathit{ph}}$ from using and FSM is only $2-5\%$ at $D/r_0=8$, as shown in Fig. \ref{fig: FSM_Dr0vsSR}. However, an upstream FSM becomes critical for designs with longer focal lengths.
%Next, we analyzed the effects of lower-order mode correction upstream of the WFC. We used an FSM to correct these lower-order modes, and with a short ($\sim 1,mm$) MLA, the focal spots were confined within the surface of the grating coupler. The FSM is therefore crucial for the WFC with the grating coupler model used in the WFC. Without an FSM, approximately one-third of the focal spots would still be coupled but effectively decreasing the WFC's coupling efficiency. Despite the reduced efficiency, we examined the correction performance of the WFC. Figure \ref{fig: FSM_Dr0vsSR} compares this performance with and without mathematical lower-order correction from the WF. In the worst-case scenario of turbulence strength ($D/r_0=8$), the photonic Strehl ratio drops by approximately 3-6\%, depending on the MLA configuration in the WFC. It is important to note that this drop is solely due to the lack of lower-order mode correction by the WFC and does not account for light lost from focal spot shifts due to tip/tilt. Commonly used DMs face a trade-off between implementing tip/tilt correction upstream and the size of the DM, owing to the DM's limited stroke.

The effect of scintillation on the photonic WFC was extensively examined in \cite{Patel:24}. The mismatch in field amplitude between the sampled subapertures cannot be corrected by the chip, so the combination loss it causes needs to be carefully characterized. 
%Finally, we examined the effects of scintillation on our photonic WFC. The co-phased beam in each channel can experience amplitude mismatch due to scintillation. Since we cannot mitigate this amplitude mismatch, it is necessary to characterize the combining loss from this mismatch by studying its effects on the photonic Strehl ratio. In \cite{Patel:24}, we analyzed this for a single configuration of 61 hexagons. 
The mismatch depends on how the wavefront is sampled, i.e. the size and configuration of the MLA. The closed-form Eq. \ref{eq: SR_ph_fiterr} relates $\mathit{SR}_{\mathit{ph}}$ to the size of the MLA and the scintillation index. This empirical relation is used to characterize the performance of the WFC, as shown in Fig. \ref{fig: Dr0vsnumHexvsP}. It provides an estimate of the required sampling size (i.e., the MLA size) for varying turbulence strengths, which influence the scintillation introduced into the wavefront. The dependence of the scintillation coefficient ($s$) on $D/r_0$ is also provided here (see Fig. \ref{fig: SI_SC_Dr0}b) to complement the results in \cite{Patel:24}.

\section{Conclusions and future work} \label{sec: concl}
%- point out that end-to-end sims were performed and main results were already reported. The goal of this complementary work is to study how three artifacts/effects/improvements affect performance.
An end-to-end simulation of the WFC discussed here was presented in \cite{Patel:24}, and the details of an experimental demonstration are provided in \cite{diab2024experimental}. The simulation results quantify the performance of the corrector under varying atmospheric turbulence conditions. This work examined three secondary effects that might also impact the device's performance: the obscuration of the telescope pupil (Sec. \ref{sec: pupilplaneMLA}), tip/tilt pre-correction with an FSM (Sec. \ref{sec: FSM}), and scintillation effects (Sec. \ref{sec: scinti}). None of these three effects were found to limit the performance of the photonic WFC. To validate these simulation results, the AO bench described in \cite{diab2024experimental} will need to be upgraded to include masks for the aperture functions, an upstream FSM, and additional phase screens with appropriate propagation distances between them.   
%We presented a photonic-based WFC designed for efficient single-mode fiber coupling. An end-to-end simulation of the WFC to estimate the device's internal efficiency is detailed in \cite{Patel:24}, and an experimental demonstration of the concept is provided in \cite{diab2024experimental}. This study serves as a complementary analysis to estimate the performance of the WFC under external factors, including the secondary obscuration of the telescope pupil, tip/tilt correction, and scintillation effects in the wavefront.

%We demonstrated that the WFC requires no additional mitigation when used with a telescope that has secondary obscuration, with the only loss being due to reduced coupling caused by the obscuration. While the lower-order modes can be effectively corrected by the WFC, the photonic-based WFC needs an FSM to minimize coupling loss caused by shifts in the focal spots from the MLA lenslets. To further minimize this loss and potentially eliminate the need for an FSM, we could explore grating coupler designs with larger active regions and even shorter MLAs. Finally, we estimated the optimal MLA configuration needed to effectively sample the wavefront, thereby reducing amplitude mismatch and minimizing the combining loss at the MMI beam combiner and reaching the performance required from the WFC.

\acknowledgments % equivalent to \section*{ACKNOWLEDGMENTS}       
 
This work was supported by the High Throughput and Secure Networks Challenge Program of the National Research Council of Canada (HTSN 647 and 628). The authors also acknowledge CMC Microsystems for the provision of products and services that facilitated this research.

% References
\bibliography{report} % bibliography data in report.bib
\bibliographystyle{spiebib} % makes bibtex use spiebib.bst

\end{document}